\documentclass[12pt]{iopart}
\input epsf
\usepackage{iopams}

\newcommand{\apj}{\textit{Astrophys.\ J.} }
\newcommand{\apjs}{\textit{Astrophys.\ J.\ Suppl.\ Ser.} }
\newcommand{\prb}{\textit{Phys.\ Rev.} B }
\newcommand{\pre}{\textit{Phys.\ Rev.} E }
\newcommand{\pra}{\textit{Phys.\ Rev.} A }
\newcommand{\rmp}{\textit{Rev.\ Mod.\ Phys.} }

\newcommand{\mearth}{\,\mathrm{M}_\oplus}

\newcommand{\gcc}{\mathrm{~g~cm}^{-3}}

\begin{document}

\article[Dense plasmas in astrophysics]{Strongly
Coupled Coulomb Systems - 2005}{Dense plasmas in astrophysics:
from giant planets to neutron stars}

\author{G Chabrier$^1$, D Saumon$^2$ and A Y Potekhin$^3$
}

\address{$^1$ Ecole Normale Sup\'erieure de Lyon, C.R.A.L.(UMR 5574 CNRS), France}
\address{$^2$ Los Alamos National Laboratory, NM 87545, USA}
\address{$^3$ Ioffe Physico-Technical Institute, St Petersburg, Russia}

\begin{abstract}
We briefly examine the properties of dense plasmas characteristic of the interior of giant planets and the atmospheres
of neutron stars. Special attention is devoted to the equation of state of hydrogen and helium at high
density and to the effect of magnetic fields on the properties of dense matter.

\end{abstract}

\submitto{\JPA}

\section{Introduction}
An accurate determination of the thermodynamic properties of matter under extreme conditions of temperature and
density is required for a correct description of the mechanical and thermal properties of many
dense astrophysical bodies,
including giant planets, low-mass stars (i.e., stars
smaller than the Sun), and so-called compact stars (white and brown dwarfs and neutron stars).
These objects  are composed dominantly of 
ion-electron plasmas, where ions are strongly correlated and electrons
are strongly or partially degenerate:
 classical Coulomb coupling parameter
$\Gamma_\mathrm{i}=(Z_\mathrm{i}e)^2/k_BTa_\mathrm{i}$ is
large and electron density parameter $r_s=a_\mathrm{i}/(a_0 Z_\mathrm{i}^{1/3})$ is
less than unity
(here
$a_0=\hbar^2/(m_e e^2)$ denotes the electronic Bohr radius,
 $a_\mathrm{i}=(3/4\pi n_\mathrm{i})^{1/3}$ the mean inter-ionic distance,
 $Z_\mathrm{i}$ the ion charge number, and $n_\mathrm{i}$ the ion number density).
The correct description of the structure and cooling of these astrophysical bodies thus requires the knowledge
of the equation of state (EOS) and the transport properties of such dense plasmas.
In this short review, we focus on the two extremes of this range of astrophysical objects in term of matter density:
Jovian planets and neutron stars. As will be shown in the next sections, modern experiments and observations
provide stringent constraints on the thermodynamic properties of dense matter under the physical conditions characteristic of these objects.

\section{The equation of state of hydrogen and the structure of Jovian planets}

\subsection{Hydrogen pressure dissociation and ionization}
\label{hug}

Jupiter and Saturn are composed of about 70\%--97\% by mass of hydrogen and helium. Temperatures and pressures
range from $T=165$ K and $T=135$ K at $P=1$ bar, respectively, at the surface, to $T>8000$ K, $P>10$ Mbar at the center. At pressures around
$P\sim 1$--3 Mbar, corresponding to about
$80\%$ and $60\%$ of the planet's radius, as measured from the planet center,  for Jupiter and Saturn, respectively,
hydrogen undergoes a transition from an insulating molecular phase to a conducting ionized plasma. The description of this transition, described
as the pressure-ionization or metallization of hydrogen, has remained a challenging problem since the pioneering work of Wigner \& Huntington \cite{WH35}.
Much experimental work has been devoted to this problem, but no conclusive result has been reached yet.
Several high-pressure
shock wave experiments have been conducted in order to probe the EOS of deuterium,
the isotope of hydrogen, in the regime of pressure ionization. Gas gun shock compression experiments were
generally limited to pressures below 1 Mbar \cite{Nellis83}, probing only the domain
of molecular hydrogen.
New techniques include laser-driven shock-wave experiments \cite{Collinsetal98,Mostovych}, pulse-power compression experiments \cite{Knudsonetal04} and
convergent spherical shock wave experiments \cite{Belovetal02, Boriskovetal03} and
 can achieve pressures
up to 5 Mbar in fluid deuterium at high temperature, exploring
for the first time the regime of pressure-dissociation and ionization. These recent experiments
give different results at $P\gtrsim 1$ Mbar, however, and this controversy needs to be settled before a robust comparison between
experiment and theory can be made in
 the very domain of hydrogen pressure ionization.

On the theoretical front, a lot of effort has been devoted to describing the pressure ionization of hydrogen. The EOS commonly used
for modeling Jovian planet interiors is the
Saumon-Chabrier-Van Horn (SCVH) EOS \cite{SC91, SC92, SCVH} wich includes a detailed description of the partial ionization regime. This EOS
reproduces the Hugoniot data of Nellis \etal 
\cite{Nellis83} but
yields temperatures about 30\% higher than the gas reshock data, indicating insufficient D$_2$
dissociation \cite{Holmes}. A slightly revised version \cite{Saumonetal00} recovers the gas gun reshock temperature data as well as the laser-driven shock wave
results
\cite{Collinsetal98}, with a maximum compression factor of $\rho/\rho_0\simeq 6$,
where $\rho_0=0.17\,\gcc$ is the initial density of liquid deuterium at 20 K.
On the other hand, the earlier
SESAME EOS
\cite{Kerley}, based on a similar formalism, predicts a smaller compression factor, with $\rho/\rho_0\simeq 4$,
in general agreement with all the other recent shock wave experiments. \textit{Ab initio} approaches for the description of dense hydrogen
include path integral Monte Carlo (PIMC)
 \cite{MilitzerCeperley, Militzeretal, Bez04} and
quantum Molecular Dynamics (QMD) simulations.
The latter 
combine molecular dynamics (MD) and Density Functional Theory (DFT) to take into account the quantum nature of the electrons
\cite{Lenosky, Bagnier, Desjarlais, Bonev}. The relevance of earlier MD-DFT calculations was questioned on the basis that these simulations were
unable to reproduce data from gas-gun experiments \cite{Lenosky}. This problem has been solved with more accurate simulations \cite{Bagnier, Desjarlais, Bonev}.
Although an \textit{ab initio} approach is more satisfactory than the phenomenological approach
based on effective potentials, in practice these simulations
also rely on
 approximations, such as the handling of the so-called
sign problem for the antisymmetrization of the fermion wave functions, or the calculation of the electron functional density itself
(in particular the exchange and correlation effects), or the use of effective pseudo-potentials of restricted validity, not mentioning finite size effects. Moreover, these simulations are too
computationally intensive for the calculation of
an EOS covering several orders of magnitude in density and temperature, as necessary for the description of the structure and evolution of
astrophysical bodies.

\begin{figure}
\begin{center}
\epsfxsize=125mm
\epsfbox[18 190 592 660]{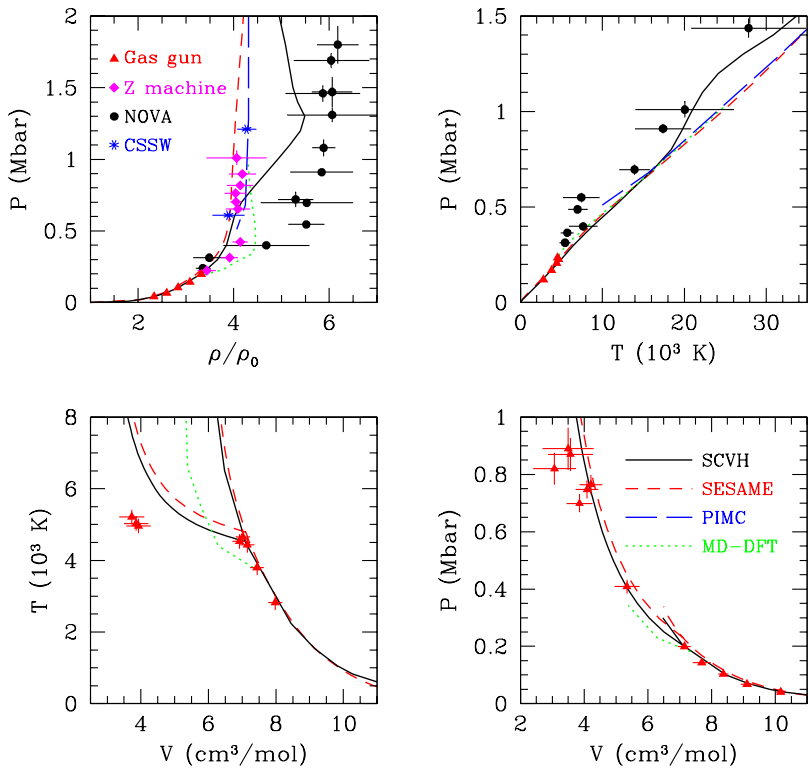}
\end{center}
\caption{
Experimental shock $(P,\rho,T)$ data and theoretical Hugoniots of deuterium. Sources of data are:
Gas gun \etal \cite{Nellis83, Holmes}, Z machine \cite{Knudsonetal04}, NOVA \cite{Collinsetal98, Collinsetal01} and CSSW \cite{Belovetal02, Boriskov}. 
Curves show Hugoniots computed from the EOSs of SCVH \cite{SCVH}, SESAME \cite{Kerley}, 
PIMC \cite{MilitzerCeperley}, and MD-DFT \cite{Desjarlais}.
}
\label{Hug}
\end{figure}

Figure \ref{Hug} compares experimental and theoretical Hugoniots in the $P$-$\rho$ and $P$-$T$ planes.
The disagreement between the laser-driven experiments and the other techniques is illustrated in
the $P$-$\rho$ diagram. Whereas the SCVH EOS achieves a maximum compression similar to the laser-driven data,
all the other models predict compression factors in the $P$-$\rho$ plane in agreement with the more recent data. The MD-DFT results, however,
predict temperatures for the second shock significantly larger than the
experimental results \cite{Holmes}. Even though the experimental double-shock temperature may be underestimated due to unquantified thermal conduction
into the window upon shock reflection, and thus represents a lower limit on the reshock temperatures, the disagreement in the $T$-$V$
plane is significant. As noted previously, the degree of molecular dissociation has a significant influence on the thermodynamic properties
of the fluid and insufficient dissociation in the simulations may result
in overestimates of the temperature. It has
been suggested that the LDA/GGA approximations used in MD-DFT underestimate the dissociation energy of D$_2$ \cite{Stadele}. This would lead to even less dissociation. The fact that compression along the experimental Hugoniot remains small thus suggests compensating effects in the case of hydrogen. More recent,
improved simulations \cite{Bonev}, however, seem to partly solve this discrepancy and to produce reshock temperatures in better
agreement with the experimental results. Peak compression in the modern MD-DFT simulations occurs in the $\sim 0.2$--0.5 Mbar
range around a dissociation fraction of $\sim 50\%$.

The differences in the behaviour of hydrogen at high density and temperature illustrated by the various results displayed
in Fig. \ref{Hug} bear important consequences
for the structure and evolution of our Jovian planets.
These differences must be correctly understood
before the description of hydrogen pressure dissociation and ionization stands on firm grounds.
As noted by Boriskov \etal \cite{Boriskov}, all the recent experiments agree quite well in terms of the shock speed $u_s$ versus the particle velocity $u_p$, almost within their respective error bars.  Error bars and differences in $(u_s,u_p)$ are amplified
in a $P$--$\rho$ diagram by a factor of  $({\rho / \rho_0}-1)$.  These are challenging experiments as the differences seen in
panel 1 of Fig.~1 arise from differences in $u_s$ and $u_p$ of less than 3\%. 

\subsection{Jupiter and Saturn interiors}

The rapid rotation of Jovian planets induces a nonspherical gravitational field that can be expanded 
in Legendre polynomials $P_n(\cos \theta)$:
\begin{eqnarray}
V(r,\theta)=-{GM\over r}\Bigl[1-\sum_{n=1}^\infty \left({R_{eq}\over r}\right)^nJ_nP_n(\cos \theta)\Bigr],
\end{eqnarray}
where $M$ and $R_{eq}$ denote respectively the planet mass and equatorial radius, 
and the $J_n$ are the gravitational moments:
\begin{eqnarray}
J_n=-{1\over MR_{eq}^n}\int_Vr^{\prime n}P_n(\cos \theta)\rho(r^\prime,\theta)\,d^3r^\prime .
\end{eqnarray}

\begin{figure}
\begin{center}
\epsfxsize=140 mm
\epsfbox[18 430 592 680]{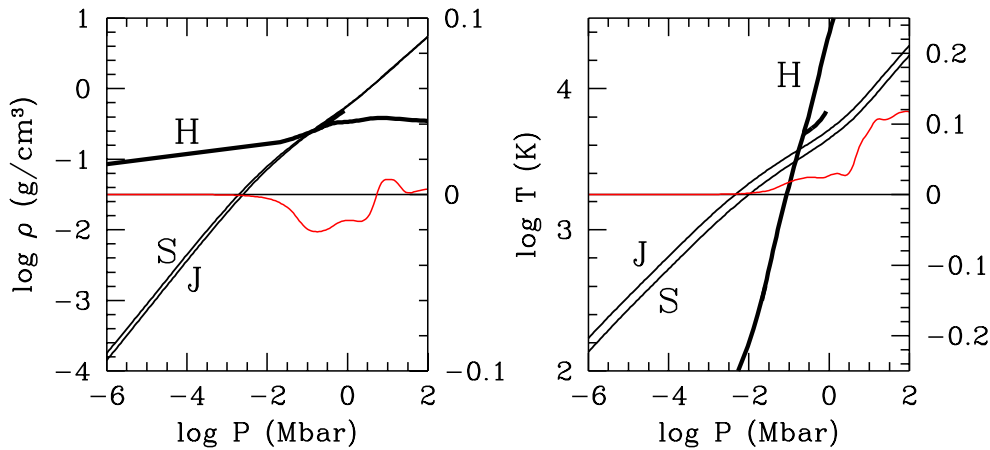}
\end{center}
\caption{Adiabats for hydrogen in $P$-$\rho$ and $P$-$T$ planes. The curves labeled "S" and "J" show the SCVH-interpolated EOS adiabats of Saturn
and Jupiter, determined by $T=136$ K and $T=170$ K at $P=1$ bar, respectively. The first and second-shock Hugoniots calculated
with the SESAME EOS are shown by the heavy solid line labeled "H". The light solid curve (right-hand scale) shows
the \textit{difference} between Jupiter adiabats calculated with the SESAME EOS relative to the SCVH-interpolated EOS.}
\label{JS-adiabat}
\end{figure}

Because of north-south symmetry, the moments of odd order are null. The first three nonvanishing moments, $J_2$, $J_4$ and $J_6$
have been measured with high accuracy for both planets during spacecraft flyby missions. Combined with the planet mass, radius
and rotation period, these provide integral constraints on the density profile  of the planet, $\rho(r,\theta)$, to be compared with the
corresponding values from a structure model
obtained for a self-gravitating and rotating fluid body in hydrostatic equilibrium. The EOS provides the $P(\rho)$
relation needed to close the system of equations. The structure of the H/He envelopes of giant planets is fixed by the specific entropy
determined from observations at their surface. The very high efficiency of convection in the interior of these objects
leads to nearly adiabatic interior profiles.
The structure of the planet is thus determined by the choice of the hydrogen and to a lesser
extent by the helium EOS used in the models.
A detailed study of the influence of the EOS of hydrogen on the structure and evolution
of Jupiter and Saturn has been conducted recently \cite{SG04}.
Fortunately, some shock wave experiments overlap
Jupiter's and Saturn's adiabats. Figure \ref{JS-adiabat} displays Jupiter (J) and Saturn (S) adiabats for hydrogen calculated
with the SCVH EOS and the first and second shock Hugoniots calculated with the SESAME EOS and illustrates \textit{relative differences}
in density between Jupiter adiabats computed with these two EOSs. As demonstrated by Saumon \& Guillot \cite{SG04}, the small
($\le 5\%$) difference on the $(P,\rho)$ relation along the adiabat between the two EOSs, representative of the two sets of
experimental results, is large enough to affect appreciably the interior structure of the models. Note that
the SESAME D$_2$ Hugoniot at low density is somewhat stiffer than the gas gun experiments \cite{Nellis83} and does not
recover the ideal H$_2$ entropy at low temperature and density. No model of Jupiter
could be obtained with this EOS \cite{SG04}. A slightly modified SESAME EOS, which does recover the H$_2$ entropy at low temperature and density,
yields Jupiter models with a very small core mass, $M_\mathrm{core}\sim 1\,\mearth$ ($\mearth$ is the mass of the Earth) and a mass $M_Z\sim 33\,\mearth$ of heavy elements ($Z>2$)
mixed in the H/He envelope. The SCVH EOS yields models with $M_\mathrm{core}\sim$0--6 $\mearth$ and $M_Z\sim 15$-26$\,\mearth$.
Models of Saturn are 
less sensitive to the EOS differences, since only $\sim 70\%$ of its mass lies at $P>1$ Mbar, compared to 91\%
for Jupiter.
Models computed with the SCVH and the modified SESAME EOS have $M_{\rm core}=10$--21$\,\mearth$ and $M _Z=20$--27$\,\mearth$
and 16--29$\,\mearth$, respectively.
As seen in Fig. \ref{JS-adiabat}, the temperature along the adiabat is more sensitive to the choice of the EOS. This affects
the thermal energy content of the planet and thus its cooling rate and evolution.
Equations of state which are adjusted to fit the deuterium reshock temperature measurements \cite{Ross98} lead to models that take
$\sim 3\,$Gyr for Jupiter to cool to its present state.
Even when considering uncertainties in the models,
or considering the possibility of a  H/He phase separation, 
such
a short cooling age is unlikely to be reconciled with the age of the solar system.
This astrophysical constraint suggests that
the reshock temperature data are too low.

\subsection{Helium equation of state. Plasma phase transition}

The planet interior models are also affected, to a lesser extent, by the uncertainties of the helium EOS. A model EOS for
helium at high density, covering the regime of pressure ionization, has been developed recently by Winisdoerffer and Chabrier
\cite{WC05}. This EOS, based on effective interaction potentials between He, He$^+$, He$^{++}$ and e$^-$ species, reproduces
adequately experimental Hugoniot and sound speed measurements up to $\sim 1$ Mbar. In this model,
pressure ionization is predicted to occur directly from He to He$^{++}$. Because of the uncertainties in the treatment of the
interactions at high density, however, the predicted ionization density ranges from a few to $\sim 10 \,\gcc$.
Comparison of the model predictions with available measurements of electrical conductivity of helium at high density \cite{Ternovoi01, Fortov03}
is under way.

The pressure ionization and metallization of hydrogen have been predicted
to occur through a first order phase transition, the so-called plasma phase transition (PPT) \cite{WH35, Norman, Ebeling, SC89, SC92, KI98}.
Nearly all of these PPT calculations are based on chemical EOS models.  Such models are based on a Helmholtz free energy that includes
contributions from 1) neutral particles (atoms and molecules), 2) a fully ionized plasma, and 3) usually a coupling between the two.  It is well-known
that realistic fully ionized plasma models become thermodynamically unstable at low temperatures and moderate densities.  This is analogous to 
the behavior of expanded metals at $T=0$ that display a region where $dP/d\rho<0$ and even $P<0$ \cite{PinesNozieres}.  This behavior of the fully ionized plasma
model is formally a first order phase transition and reflects the formation of bound states in the real system.  In other words, the
chemical models have a first order phase transition built in from the onset, and this phase transition coincides, not surprisingly,
with the regime of pressure ionization. This represents a common flaw in this type of models and it follows that their prediction
of a PPT in hydrogen is not credible.  This is further supported by a detailed study of two of these models
\cite{SC92, KI98}.  On the other hand, recent \textit{ab initio} simulations find a sharp ($6\pm 2\%$) volume discontinuity
at constant pressure \cite{Scandalo, Bonev}  or $dP/dT<0$ at constant volume \cite{Magro, BBC, Filinov}, a feature consistent with
the existence of a first order phase transition. At the same time, the pair correlation function exhibits
a drastic change from a molecular to an atomic state with a metallic character
(finite density of electronic states at the Fermi level). These transitions are found to occur in the $\sim 0.5$--1.25 Mbar
and $\sim 1500$--3000 K temperature range.
While these results are suggestive, a systematic exploration of this part of the phase diagram remains to be done. Note that a first order structural transition for H$_2$ at $T=0$ is predicted to occur
at a pressure $P\gtrsim 4.0$ Mbar, from DFT calculations based on \textit{exact} exchange calculations \cite{Stadele}.
There is so far no published experimental evidence for the PPT but it cannot yet be ruled out.  
Given the difficulty of modeling this region of the phase diagram of hydrogen, 
only experiments can ultimately establish whether a PPT exists or not.

\section{Dense matter in strong magnetic fields. Neutron star structure and cooling}

Neutron stars (NS) consist of a nucleon core
and envelopes composed of Coulomb plasmas.
Cooling rates of these
stars are determined by the 
heat capacity and neutrino emission processes in their cores and by 
heat transport in the envelopes.
For the neutrino emission,
most important are so called direct Urca (Durca) processes (beta-decay and beta-capture)
and modified Urca (Murca) processes (the same but with participation of an additional 
nucleon, which helps to fulfill momentum conservation). 
The Murca processes are less powerful, but they work in every
sufficiently hot NS. At contrast, the most powerful Durca processes operate only if
the proton fraction in the core is large enough
(otherwise the momentum conservation condition
for degenerate nucleons cannot be satisfied). 
Some models of nuclear matter predict that a NS with relatively high
mass should have a sufficient proton fraction at the stellar center for the Durca processes to occur. 
Such stars should cool faster, which opens a possibility to check 
the EOS of superdense matter through observations.
The cooling rates are also strongly affected by nucleon superfluidity
(see \cite{YakPethick} for review and references).

Most NSs have magnetic fields $B\sim10^{11}$--$10^{13}$~G, 
whereas some (so-called magnetars)
are thought to have fields as high as $\sim10^{14}$--$10^{15}$~G.
The photosphere of a NS is characterized by temperatures
$T_\mathrm{s}\simeq 10^5$--$10^7$ K (depending on the age $t$ and mass $M$ of the star)
and densities
$\rho\simeq 10^{-2}$--$10^4\,\gcc$ (depending on $T$ and $B$).
Traditionally the NS crust is assumed to be composed of iron. However, the outer layers,
including the atmosphere, can be composed of light elements (H, He, C) 
accreted on top of the iron layer. Therefore the determination of
the temperature profiles and emitted spectra of NSs
 requires an accurate description of the formation of 
bound states and pressure ionization of these elements
in a strong magnetic field.

  The quantum-mechanical properties of free charged particles
and bound species (hydrogen atoms and molecules) are strongly modified by
the magnetic field, which thereby affects the thermodynamic properties of the plasma \cite{VP01, Lai01}.
 The transverse motion of electrons in a magnetic field
is quantized into Landau levels. The energy of the $n$th
Landau level of the electron (without the rest energy) is
$m_e c^2(\sqrt{1+2bn}-1)$,
which becomes $\hbar\omega_\mathrm{c} n$ in the non-relativistic limit,
where
$
\hbar \omega_\mathrm{c}= \hbar {eB/ m_e c} = 11.577 B_{12} \,\mathrm{keV},
$
is the electron cyclotron energy,
$
b=\hbar\omega_\mathrm{c}/m_e c^2=B_{12}/44.14
$
is the field strength in the relativistic units,
and $B_{12}=B/(10^{12}\,\mathrm{G})$ is a typical
magnetic-field scale for NS conditions.
The atomic unit for the magnetic-field strength is set by $\hbar \omega_\mathrm{c}=e^2/a_0$,
i.e., $B_0=(m_ec/\hbar e)\times(e^2/a_0)=2.35\times 10^9$ G.
It is convenient to define a dimensionless magnetic-field strength
$
\gamma= {B/ B_0}=b/\alpha_f^2,
$
where $\alpha_f$ is the fine structure constant.

 For $\gamma\gg 1$, as encountered in NSs,
the ground-state atomic and molecular binding energies increase
as $\sim\ln^2\gamma$.
The H atom in a strong magnetic field is compressed in the transverse directions to the radius $\sim a_\mathrm{m}$, where
\begin{equation}
a_\mathrm{m}= ({\hbar c / e B})^{1/2}= \gamma^{-1/2} a_0 = 2.56\times 10^{-10} B_{12}^{-1/2} \,\mathrm{cm}
\label{eqn_cyclorad}
\end{equation}
is the quantum magnetic length, which
becomes the natural length unit. 
The increase of binding energies and decrease of sizes
lead to a significant increase of the fraction of non-ionized atoms 
in the plasma at the photospheric densities (which are the higher
the stronger the magnetic field). 
For example, at $T=10^6$~K and $B=10^{13}$~G,
the typical density is $\rho\sim1$ g cm$^{-3}$, and
there are $>1$\% of atoms in the H atmosphere.
Because of the alignment of the electron spins antiparallel to the field, two atoms in their ground state
($m=0$) do not bind together, because of the Pauli exclusion principle. 
One of the two H atoms has to be excited in the $m=-1$ state to form the ground state of the
H$_2$ molecule \cite{Lai01}.
Another important effect is that thermal motion of atoms 
across the field strongly modifies their 
binding energies and radiative transition rates.
As shown in \cite{PCS99,PC03,PC04},
the allowance for partial ionization and thermal motion 
is crucial for
neutron-star atmosphere modelling. 

As long as $T\ll \hbar\omega_\mathrm{c}/k_\mathrm{B} = 1.343\times10^8\,B_{12}$~K and
$\rho \ll \rho_B \approx 7.1\times10^3\,B_{12}^{3/2}\gcc$,
the electron cyclotron energy $\hbar \omega_\mathrm{c}$ exceeds both the thermal energy
$k_\mathrm{B}T$ and the electron Fermi energy $k_\mathrm{B} T_\mathrm{F}$,
so that 
the field is \textit{strongly quantizing} (e.g., \cite{VP01}).
In this case, typical for the NS photospheres,
the electron spins are aligned antiparallel to the field.
The electron Fermi energy decreases, therefore the onset of degeneracy
is shifted to higher densities (slightly below $\rho_B$).
Proton motion is also quantized by the magnetic field,
but the corresponding cyclotron energy is much smaller,
$\hbar\omega_\mathrm{cp}= \hbar\omega_\mathrm{c} m_e/m_p$.

A model which describes the
thermodynamics
of an interacting (H$_2$, H, H$^+$, $e^-$) plasma in a strong magnetic field was constructed in \cite{PCS99}.
On the base of this model, the EOS for magnetized H atmospheres of NSs, as well as 
their opacities,
were explored and tabulated in \cite{PC03,PC04}.

Landau quantization of electron orbits affects not only the EOS and the radiative
opacities,
but also the heat conduction in the surface layers
(see \cite{VP01} and references therein).
The EOS of strongly magnetized, partially ionized hydrogen plasma as well as
the electron conductivities 
and radiative opacities 
in neutron star magnetized envelopes
were used in \cite{P03} to calculate the thermal structure and
cooling of superfluid NSs
with accreted envelopes in the presence of strong dipole magnetic fields. 
In \cite{P03} (see also \cite{Yak05} and references therein), 
the effect of neutron superfluidity
in the NS inner crust was also examined.
The account of
the effects of
 accreted matter, magnetic field and neutron
superfluidity alters
significantly 
the NS cooling.

\begin{figure}
\begin{center}
\epsfxsize=150mm
\epsfbox{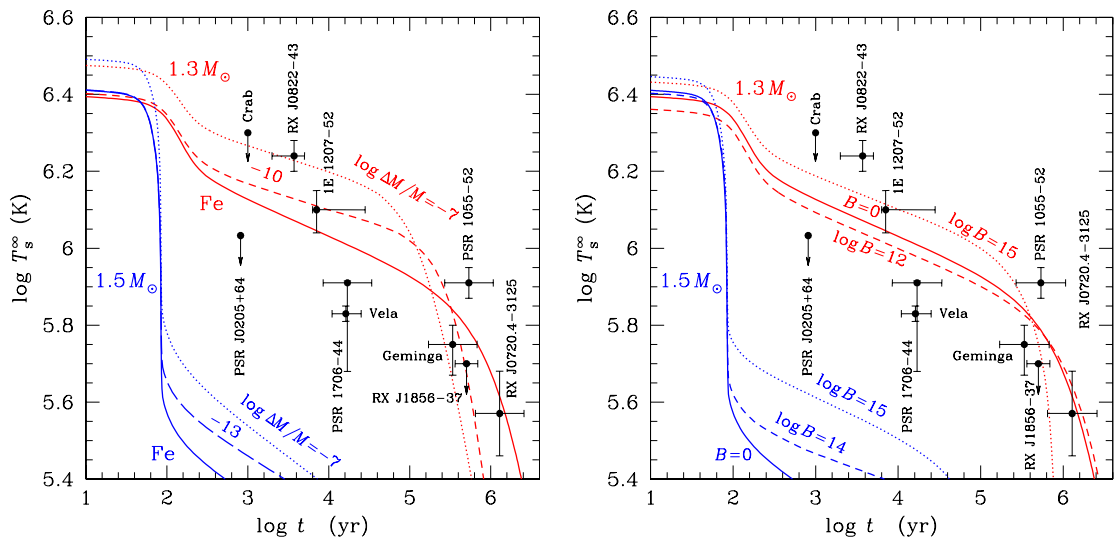}
\end{center}
\caption{Effective surface temperature (as seen by distant observer, $T_\mathrm{s}^\infty$) versus
NS age $t$ for assumed NS mass $M=1.3$ and 1.5 solar masses.
The dots with error bars show estimates of NS ages and effective temperatures from 
various observations; the dots with errors indicate observational upper limits.
\textit{Left}: Cooling of NSs with different relative masses $\Delta M/M$
of accreted (H-He-C) matter (values of $\log\Delta M/M$ are indicated near the curves)
Solid curves refer to non-accreted (Fe) iron envelope of the star.
\textit{Right}: Cooling of NSs with iron envelope for different
magnetic field strengths ($\log B$ in Gauss).}
\label{cooling_NS}
\end{figure}

Figure \ref{cooling_NS} displays theoretical cooling curves of NSs with
(lower curves, $M=1.5\,M_\odot$) or without (upper curves, $M=1.3\,M_\odot$)
Durca processes in the core, with or without accreted envelopes, and
with magnetic field of different strengths,
compared to
the estimates of the effective temperature obtained from
 observations (see \cite{YakPethick} for references).
As seen in the figure, the presence of a light-element (accreted) envelope
increases $T_\mathrm{s}$ at the early cooling stage ($t\lesssim10^5$~yr),
and as a result the thermal energy becomes exhausted sooner. 
The magnetar-like magnetic field $B\gtrsim10^{14}$~G acts in a similar way, 
whereas a weaker field almost does not affect the cooling.

For simplicity,
in Fig.~\ref{cooling_NS} we neglect the effects of superfluidity. Their discussion can be found in \cite{YakPethick,P03,Yak05}.

\section{Conclusions}

In this brief review, we considered the description of the thermodynamic properties of dense Coulomb matter in two specific astrophysical
contexts, Jovian planets and neutron stars. The description of the pressure ionization of hydrogen and other elements,
as well as the presence of strong magnetic fields, play an important role in determining the mechanical and thermal properties and the evolution of these objects.
Models including these complex effects can successfully explain a variety of observations.
On the other hand, modern experiments and/or observations can enable
us to discriminate between various EOS models 
in planet interiors and lead to a better determination of masses of accreted envelopes,
surface magnetic fields
and eventually the EOS of superdense matter in neutron stars.

\ack
We thank D.\,G.~Yakovlev for providing us with his compilation of
observational data and performing
cooling calculations with our physics input for Fig.~\ref{cooling_NS}.
The work of GC was partially supported by the CNRS French-Russian grant PICS 3202. 
The work of AYP was partially supported by the RLSS grant 1115.2003.2 and the RFBR grants 05-02-16245, 03-07-90200 and 05-02-22003.
The work of DS was supported in part by the United States Department of Energy under contract W-7405-ENG-36.

\end{document}